\documentclass[twoside,english,sort&compress]{iopart}
\usepackage[T1]{fontenc}
\usepackage[latin9]{inputenc}
\usepackage{geometry}
\usepackage{units}
\usepackage{amstext}
\usepackage{graphicx}
\usepackage{esint}
\usepackage[numbers]{natbib}

\makeatletter

\providecommand{\tabularnewline}{\\}

\usepackage{iopams}
\usepackage{setstack}

\newcommand{\eqref}[1]{(\ref{#1})}

\makeatother

\usepackage{babel}
\begin{document}
\title[Spontaneous emission of a sodium Rydberg atom close to an optical
nanofibre]{Spontaneous emission of a sodium Rydberg atom close to an optical
nanofibre}
\author{E Stourm$^{1}$, Y Zhang$^{2}$, M Lepers$^{1,3}$, R Guérout$^{4}$,
J Robert$^{1}$, S Nic Chormaic$^{5}$, K Mølmer$^{2}$ and E Brion$^{1,6}$ }
\address{$^{1}$Laboratoire Aimé Cotton, CNRS, Université Paris Sud, ENS Paris
Saclay, CNRS, Université Paris-Saclay, 91405 Orsay, France.\\$^{2}$Department
of Physics and Astronomy, Aarhus University, Ny Munkegade 120, DK-8000
Aarhus C, Denmark.\\$^{3}$Laboratoire Interdisciplinaire Carnot
de Bourgogne, CNRS, Université de Bourgogne Franche-Comté, 21078 Dijon,
France.\\$^{4}$Laboratoire Kastler Brossel, UPMC-Sorbonne Universités,
CNRS, ENS-PSL Research University, Collège de France, Campus Jussieu,
F-75252 Paris, France.\\$^{5}$Light - Matter Interactions Unit,
Okinawa Institute of Science and Technology Graduate University, Onna,
Okinawa, 904 - 0495, Japan.\\$^{6}$Laboratoire Collisions Agrégats
Réactivité, IRSAMC \& UMR5589 du CNRS, Université de Toulouse III
Paul Sabatier, F-31062 Toulouse Cedex 09, France.}
\ead{etienne.brion@irsamc.ups-tlse.fr}
\begin{abstract}
We report on numerical calculations of the spontaneous emission rate
of a Rydberg-excited sodium atom in the vicinity of an optical nanofibre.
In particular, we study how this rate varies with the distance of
the atom to the fibre, the fibre's radius, the symmetry $s$ or $p$
of the Rydberg state as well as its principal quantum number. We find
that a fraction of the spontaneously emitted light can be captured
and guided along the fibre. This suggests that such a setup could
be used for networking atomic ensembles, manipulated in a collective
way due to the Rydberg blockade phenomenon.
\end{abstract}
\noindent{\it Keywords\/}: {Rydberg atoms, Optical nanofibres, Spontaneous emission rates\\
}
\submitto{\JPB }
\maketitle

\section{Introduction}

Within the last two decades, the strong dipole-dipole interaction
experienced by two neighbouring Rydberg-excited atoms \citep{G94}
has become the main ingredient for many of the atomic quantum information
protocol proposals (see \citep{SWM10} and references therein). In
particular, this interaction can be so large as to even forbid the
simultaneous resonant excitation of two atoms if their separation
is less than a specific distance, called the blockade radius, which
typically depends on the intensity of the laser excitation and the
interaction between the Rydberg atoms. The discovery of this \textquotedblleft Rydberg
blockade\textquotedblright{} phenomenon \citep{LFC01,TFS04,SRA04,CRB05,AVG98,VVZ06}
paved the way for a new encoding scheme using atomic ensembles as
collective quantum registers \citep{LFC01,BMS07,BMM07,BPS08} and
repeaters \citep{BCA12,ZMH10,HHH10}. In this novel framework, information
is stored in collective spin-wave-like symmetric states, which contain
fully delocalized atomic excitations. Qubits are more easily manipulated
and more robust in this collective approach than in the usual single-particle
paradigm.

Scalability is one of the crucial requirements for quantum devices
and interfacing atomic ensembles into a quantum network is a possible
way to reach this goal. Photons naturally appear as ideal information
carriers and the photon-based protocols considered so far include
free-space \citep{PM09}, or guided propagation through optical fibres
\citep{BCA12}. The former has the advantage of being relatively easy
to implement, but presents the drawback of strong losses. The latter
requires a cavity quantum electrodynamics setup, which is experimentally
more involved. An alternative option would be to resort to optical
nanofibres. Such fibres have recently received much attention \citep{SGH17,NGN16}
because the coupling to the evanescent (resp. guided) modes of a nanofibre
allows for easy-to-implement atom trapping \citep{BHK04,KBH04} (resp.
detection \citep{NMM07}). This coupling increases in strength as
the fibre diameter reduces and the atoms approach the fibre surface.
It was also even shown that energy could be exchanged between two
distant atoms via the guided modes of the fibre \citep{KDN05}. This
strongly suggests that optical nanofibres could play the role of a
communication channel between the nodes of an atomic quantum network
consisting of Rydberg-excited atomic ensembles. 

In this article, we make a first step towards this goal and investigate
the emission rate of a highly-excited (Rydberg) sodium atom in the
neighbourhood of an optical nanofibre made of silica. In the perspective
of building a quantum network, we are particularly interested in quantifying
how much spontaneously emitted light can be captured and guided along
the fibre. Here, we study the influence of the atom to fibre distance,
the radius of the fibre, and the symmetry of the Rydberg state, on
the emission rates into the guided and radiative fibre modes. We find
that up to $\approx13\%$, of the spontaneously emitted light can
be captured and guided along both directions of the fibre, which is
comparable with the ratio of $\approx30\%$ obtained with a cesium
atom initially in its lowest excited state $6P_{\nicefrac{3}{2}}$
and located on the surface of a 200-nm-diameter nanofibre \citep{KDB05}.
Although the theoretical framework we use here is the same, numerical
calculations are more complex than in \citep{KDB05} due to the larger
number of transitions considered. Contrary to Ref. \citep{KDB05},
we do not take into account the atomic hyperfine structure in the
excited state, which is very small for Rydberg states \citep{AMV77}.

The article is organized as follows. In Sec. \ref{SECGeneral} we
briefly present the system and introduce the expressions of the spontaneous
emission rates. In Sec. \ref{SECNumericalResults}, we present the
results of our numerical calculations and discuss the different behaviours
observed when the atom is initially in an $s$ or $p$ Rydberg state.
Finally, in Sec. \ref{SECConclusion}, we conclude and give perspectives
of our work. \ref{APPModesGuides} and \ref{APPModesRadiatifs} provide
details about the guided and radiative electromagnetic modes, \ref{APPRates}
sketches the derivation of the spontaneous emission rates of the atom
in the presence of the nanofibre and \ref{APPAtomicData} displays
the atomic data we used in our calculations. 

\section{The system \label{SECGeneral}}

\begin{figure}
\begin{centering}
\includegraphics[width=9cm]{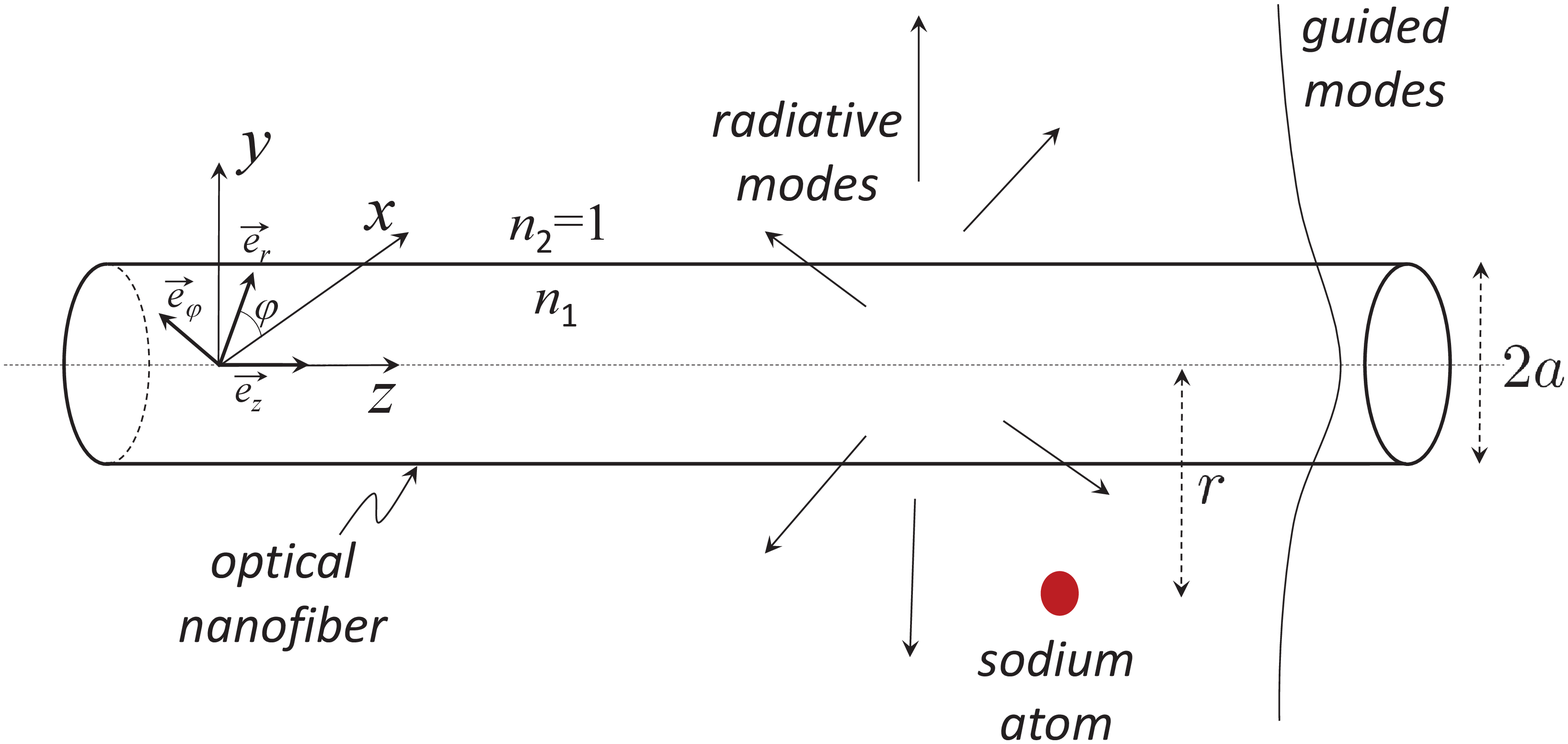}
\par\end{centering}
\caption{Sodium atom in the vicinity of an optical nanofibre with a radius
$a$. The index is $n_{1}=1.45$ for silica and $n_{2}=1$ for vacuum.
The axis of the nanofibre is arbitrarily chosen as the $z$-axis.
The cylindrical coordinates $\left(r,\varphi,z\right)$ and frame
$\left(\vec{e}_{r},\vec{e}_{\varphi},\vec{e}_{z}\right)$ are introduced.}
\label{System}
\end{figure}

We consider a sodium atom, initially prepared in the highly-excited
(Rydberg) level $n\leq10$, in the vicinity of a silica nanofibre,
whose radius is denoted by $a$ and whose axis is conventionally taken
as the $z$-axis (see figure \ref{System}). Our goal is to investigate
how the presence of the fibre modifies the spontaneous emission rate
of the atom : in particular, we want to study the influence of the
radius of the fibre, the distance of the atom to the fibre as well
as the symmetry of the Rydberg state $\left|nl_{j},m_{j}\right\rangle $
considered and the principal quantum number $n$ on the spontaneous
emission rate. Note that, though the configuration is the same as
in \citep{KDB05}, in this work, the atom is (relatively) highly excited
and, in contrast to \citep{KDB05}, several transition frequencies
must therefore be considered which complicates the numerical work.
The choice of the sodium atom and the maximal principal quantum number
$n_{\text{max}}=10$ is motivated by the fact that, for the relevant
transitions $10\rightarrow n=3,\cdots,9$, the fibre can be approximately
considered as a nonabsorbing medium of respective refractive indices
$n_{1}=\left(1.467,1.450,1.438,1.399,1.112,1.615,2.021\right)$ \citep{Pal98}.
Such constraints may, however, be alleviated by resorting to the formalism
of macroscopic quantum electrodynamics and the Green's function approach
\citep{Buh12}. These techniques allow to take the absorption of the
medium into account and therefore to deal with higher Rydberg states.
This formalism and its application to the calculation of energy shifts
will be investigated in a future work. Moreover, the choice of sodium,
rather than rubidium or cesium which are more commonly used in nanofibre
experiments, was made to allow us to neglect relativistic effects
on the electronic wavefunctions and therefore simplify our treatment.
The case of cesium will also be tackled in a future work. 

As recalled in \citep{KDB05}, the free electromagnetic field in the
presence of a cylindrical fibre can be decomposed into \emph{guided}
and \emph{radiative} modes which respectively correspond to energy
propagation along the fibre and radially to it (see \ref{APPModesGuides}
and \ref{APPModesRadiatifs}). 

\emph{Guided modes }are characterized by their frequency $\omega>0$
and order $m$, which is a positive integer fixing the periodicity
of the field with respect to $\varphi$. Due to the continuity conditions
at the core-cladding interface of the fibre, the norm of the projection
of the wavevector onto the $z$ axis, denoted by $\beta$, can only
take a discrete set of values which are the solutions of the so-called
characteristic equation equation (\ref{EQCaracteristique}) \citep{M89,SL83}.
The corresponding modes have different cutoff frequencies. In particular,
if $\omega$ is sufficiently low, only the (so-called ``hybrid'')
mode $HE_{11}$, corresponding to $m=1$, can propagate along the
fibre. Since a given mode can propagate either in the positive or
negative $z$-direction, an extra index $f=\pm1$ is introduced, such
that $\beta\times f$ is the (algebraic) projection of the wavevector
onto the $z$-axis. To complete the description, one also allows for
two different polarization directions labelled by $p=\pm1$. For the
sake of simplicity, we shall gather the characteristic numbers $\left(\omega,m,f,p\right)$
into one symbol $\mu$ and replace the discrete/continuous sums $\sum_{mfp}\int_{0}^{\infty}d\omega$
by $\sum_{\mu}$. Finally the general form of the quantized guided
field component is
\begin{eqnarray*}
\vec{E}_{g}\left(\vec{r}\right) & = & \mathrm{i}\sum_{\mu}\;\sqrt{\frac{\hbar\omega\beta'}{4\pi\epsilon_{0}}}a_{\mu}\vec{e}^{\left(\mu\right)}\left(r,\varphi\right)e^{\mathrm{i}\left(f\beta z+p\varphi\right)}+\mbox{h.c.}
\end{eqnarray*}
In this expression, $\beta'$ stands for the derivative $\left(\frac{d\beta}{d\omega}\right)$,
$\vec{e}_{\mu}$ is the electric-field profile function of the mode
$\left(\mu\right)$ whose expression is given in \ref{APPModesGuides},
while $a_{\mu}$ is the annihilation operator of the mode, satisfying
the bosonic commutation rules $\left[a_{\mu},a_{\mu'}^{\dagger}\right]=\delta\left(\omega-\omega'\right)\delta_{mm'}\delta_{ff'}\delta_{pp'}$. 

\emph{Radiative modes} are characterized by their frequency $\omega>0$,
their (positive integer) order $m$ and the projection of the wavevector
on the nanofibre axis $\beta$ which can now vary continuously between
$-\omega n_{2}/c$ and $\omega n_{2}/c$. Here, the negative or positive
sign indicates the direction of the propagation of the radiation mode
along the $z$-axis. A last number is needed to fully determine a
radiative mode, i.e. the polarization number $p=\pm1$. The two values
of $p$ correspond to two modes of orthogonal polarizations, see \ref{APPModesRadiatifs}.
For the sake of simplicity, we shall gather the characteristic numbers
$\left(\omega,\beta,m,p\right)$ into one symbol $\nu$ and replace
the discrete/continuous sums $\sum_{mp}\int_{0}^{\infty}d\omega\int_{-kn_{2}}^{kn_{2}}d\beta$
by $\sum_{\nu}$. The general form of the quantized radiative field
component is
\[
\vec{E}_{r}\left(\vec{r}\right)=\mathrm{i}\sum_{\nu}\;\sqrt{\frac{\hbar\omega}{4\pi\epsilon_{0}}}a_{\nu}\vec{e}^{\left(\nu\right)}\left(r,\varphi\right)e^{\mathrm{i}\left(\beta z+m\varphi\right)}+\mbox{h.c.}
\]
In this expression, $\vec{e}_{\nu}$ is the electric-field profile
function of the mode $\left(\nu\right)$ whose expression is given
in \ref{APPModesRadiatifs}, while $a_{\nu}$ is the annihilation
operator of the mode, satisfying the bosonic commutation rules $\left[a_{\nu},a_{\nu'}^{\dagger}\right]=\delta\left(\omega-\omega'\right)\delta_{mm'}\delta_{pp'}$. 

In the presence of the nanofibre, the spontaneous emission rate $\Gamma_{M}$
of an atom from a state $\left|M\right\rangle $ is the sum of the
rates $\gamma_{MN}$ from $\left|M\right\rangle $ to all lower states
$\left|N\right\rangle $, i.e. $\Gamma_{M}\equiv\sum_{N<M}\gamma_{MN}$
with
\begin{equation}
\gamma_{MN}\equiv2\pi\sum_{\lambda}\left|G_{\lambda MN}\right|^{2}\delta\left(\omega_{\lambda}-\omega_{MN}\right).\label{EQDecayRate}
\end{equation}
In the expression above, the sum is performed over all electromagnetic
modes denoted by $\lambda$, whether they be guided $\left(\lambda=\mu\right)$
or radiative $\left(\lambda=\nu\right)$ ; we moreover introduce the
quantities
\begin{eqnarray*}
G_{\mu MN} & \equiv & -\sqrt{\frac{\omega\beta'}{4\pi\epsilon_{0}\hbar}}\left(\vec{d}_{MN}\cdot\vec{e}^{\left(\mu\right)}\right)e^{\mathrm{i}\left(f\beta z+p\varphi\right)},\\
G_{\nu MN} & \equiv & -\sqrt{\frac{\omega}{4\pi\epsilon_{0}\hbar}}\left(\vec{d}_{MN}\cdot\vec{e}^{\left(\nu\right)}\right)e^{\mathrm{i}\left(\beta z+m\varphi\right)},
\end{eqnarray*}
characterizing the coupling of the different electromagnetic modes
to the atomic transition $\left|M\right\rangle \rightarrow\left|N\right\rangle $
of frequency $\omega_{MN}\equiv\nicefrac{\left(E_{M}-E_{N}\right)}{\hbar}$
and dipole matrix element $\vec{d}_{MN}$. Finally the decoherence
rate between states $\left|M\right\rangle $ and $\left|N\right\rangle $
is given by
\begin{equation}
\Gamma_{MN}\equiv\frac{1}{2}\left(\Gamma_{M}+\Gamma_{N}\right).\label{EQSDecoherenceRate}
\end{equation}
For a detailed derivation of equations (\ref{EQDecayRate},\ref{EQSDecoherenceRate}),
see \ref{APPRates}.

\section{Numerical results and discussion \label{SECNumericalResults}}

In this section, we present the numerical results we obtained for
the spontaneous emission rate of a sodium atom $\left(Z=11\right)$
initially prepared either in $\left|ns_{\nicefrac{1}{2}},m_{j}\right\rangle $
or $\left|np_{j},m_{j}\right\rangle $ states with $n\leq10$ and
$j=1/2$ or $3/2$. We study the influence of the principal quantum
number, $n$, and the distance from the atom to the fibre surface
on the emission rate. We also show how the fibre's radius modifies
the relative weights of the different transitions' contributions to
the total rate. For simplicity, we consider the contributions of the
guided and radiative modes separately. The atomic data we used can
be found in \ref{APPAtomicData}.

\subsection{Guided modes}

Figure \ref{FigModGuidCompsp} displays the spontaneous emission rates,
$\Gamma_{g}^{10s}$ and $\Gamma_{g}^{10p}$, of an atom initially
prepared in the states $\left|10s_{\nicefrac{1}{2}},m_{j}\right\rangle $
and $\left|10p_{j},m_{j}\right\rangle $ with $j=1/2$ or $3/2$ ,
respectively, into the \emph{guided} modes as a function of the distance
$r$ to the fibre axis (see figure \ref{System}). Note that the rates
are presented relative to the spontaneous emission rates $\Gamma_{0}^{10s}$,
$\Gamma_{0}^{10p}$ in vacuum and $r$ is expressed in units of the
fibre radius with $a=100$ nm. As expected, in both cases, the influence
of the guided modes vanishes as $r$ increases, and therefore $\Gamma_{g}^{10s},\Gamma_{g}^{10p}\rightarrow0$
when $r\rightarrow+\infty$. The maximal value is obtained for $r=a$,
i.e. when the atom is on the fibre surface. More precisely, we have
$\Gamma_{g}^{10s}\approx0.18\Gamma_{0}^{10s}$ for an atom initially
prepared in $\left|10s_{\nicefrac{1}{2}},m_{j}=\pm\frac{1}{2}\right\rangle $
and $\Gamma_{g}^{10p}\approx\left(0.027,0.035,0.044\right)\times\Gamma_{0}^{10p}$
for an atom initially prepared in $\left(\left|10p_{\nicefrac{3}{2}},m_{j}=\pm\frac{1}{2}\right\rangle ,\left|10p_{\nicefrac{1}{2}},m_{j}=\pm\frac{1}{2}\right\rangle ,\left|10p_{\nicefrac{3}{2}},m_{j}=\pm\frac{3}{2}\right\rangle \right)$.
In these calculations we assumed that the electronic wave-function
of the Rydberg atom is not affected by the nanofibre, which deserves
further study. As a more realistic configuration, we shall consider
that the Rydberg atom is located at a distance from the fibre surface
which is much larger than its radius $r_{Na}\approx5\text{ nm}=\frac{a}{20}$.
For $r=a+10r_{Na}\approx1.5\times a$, we obtain the spontaneous rate
$\Gamma_{g}^{10s}\approx0.066\times\Gamma_{0}^{10s}$ for an atom
initially prepared in $\left|10s_{\nicefrac{1}{2}},m_{j}=\pm\frac{1}{2}\right\rangle $
and $\Gamma_{g}^{10p}\approx\left(0.006,0.011,0.015\right)\times\Gamma_{0}^{10p}$
for an atom initially prepared in $\left(\left|10p_{\nicefrac{3}{2}},m_{j}=\pm\frac{1}{2}\right\rangle ,\left|10p_{\nicefrac{1}{2}},m_{j}=\pm\frac{1}{2}\right\rangle ,\left|10p_{\nicefrac{3}{2}},m_{j}=\pm\frac{3}{2}\right\rangle \right)$.
Moreover, we note that in general, $\Gamma_{g}^{10p}\ll\Gamma_{g}^{10s}$,
and $\Gamma_{g}^{10p_{\nicefrac{3}{2}},m_{j}=\pm\frac{1}{2}}<\Gamma_{g}^{10p_{\nicefrac{1}{2}}}<\Gamma_{g}^{10p_{\nicefrac{3}{2}},m_{j}=\pm\frac{3}{2}}$.
The latter relation can be qualitatively understood by geometric arguments
on the coupling of guided modes with the atomic orbitals. The more
a state is polarized along $z$, the less it couples to the guided
modes which are essentially polarized orthogonally to the fibre axis.
This is consistent with what we observe, since the states $\left|10p_{\nicefrac{3}{2}},m_{j}=\pm\frac{1}{2}\right\rangle $
are better aligned along $z$ than the states $\left|10p_{\nicefrac{1}{2}},m_{j}=\pm\frac{1}{2}\right\rangle $
which themselves are more aligned along $z$ than $\left|10p_{\nicefrac{3}{2}},m_{j}=\pm\frac{3}{2}\right\rangle $.
This can be seen on their relation with the decoupled basis states
\citep{DecoupledBasis}.
\noindent \begin{center}
\begin{figure*}
\begin{centering}
\includegraphics[width=12cm]{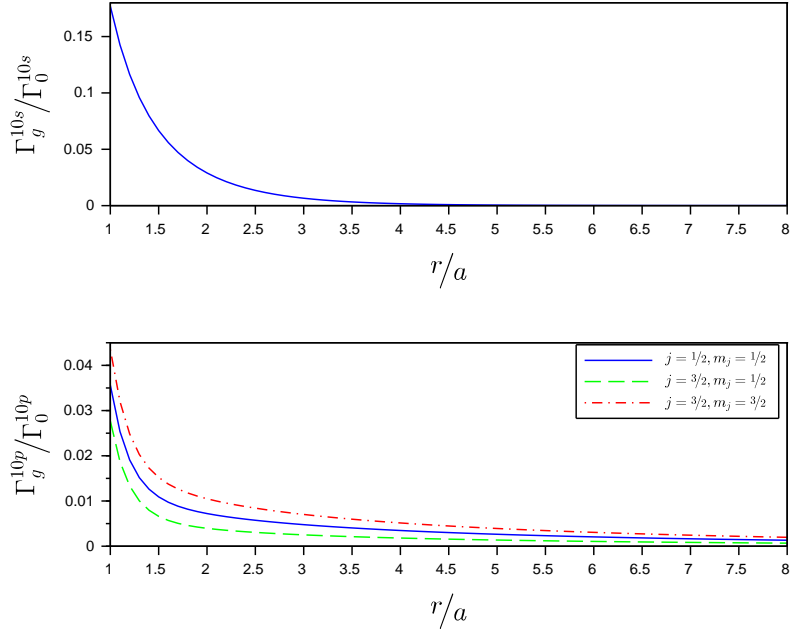}
\par\end{centering}
\caption{Spontaneous emission rate of a sodium atom into the guided modes of
a nanofibre of radius $a=100$nm. The rate is plotted as a function
of the distance $r$ of the atom to the fibre axis : (top) atom initially
prepared in the state $\left|10s_{\nicefrac{1}{2}},m_{j}=\pm\frac{1}{2}\right\rangle $,
(bottom) atom initially prepared in the states $\left|10p_{\nicefrac{1}{2}},m_{j}=\pm\frac{1}{2}\right\rangle $,
$\left|10p_{\nicefrac{3}{2}},m_{j}=\pm\frac{1}{2}\right\rangle $
and $\left|10p_{\nicefrac{3}{2}},m_{j}=\pm\frac{3}{2}\right\rangle $.
The rates $\Gamma_{g}^{10s},\Gamma_{g}^{10p}$ are presented relative
to the spontaneous emission rates $\Gamma_{0}^{10s}$, $\Gamma_{0}^{10p}$
in vacuum.}
\label{FigModGuidCompsp}
\end{figure*}
\par\end{center}

\begin{figure*}
\begin{centering}
\includegraphics[width=13cm]{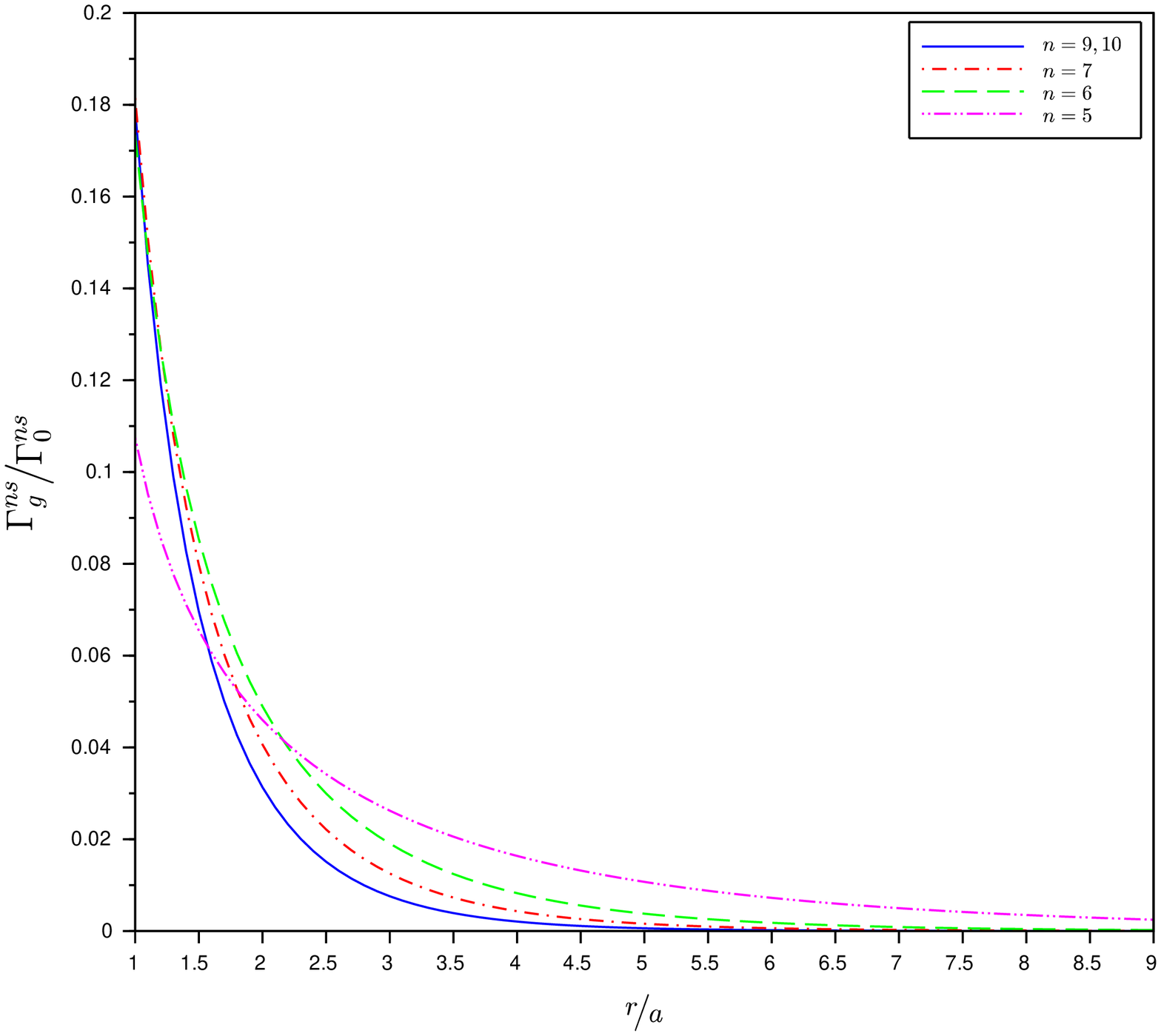}
\par\end{centering}
\caption{Spontaneous emission rate of a sodium atom initially prepared in $\left|ns_{\nicefrac{1}{2}},m_{j}=\pm\frac{1}{2}\right\rangle $,
for $n=5,\cdots,10$, into the guided modes of a nanofibre of radius
$a=100$ nm. The rate is plotted as a function of the distance $r$
of the atom from the fibre axis. The rate, $\Gamma_{g}^{ns}$, is
renormalized by the spontaneous emission rate in vacuum, $\Gamma_{0}^{ns}$,
and the distance $r$ is expressed in units of the fibre radius, $a$.}

\label{FigModGuidDepn}
\end{figure*}

Figure \ref{FigModGuidDepn} shows the influence of the principal
quantum number $n$ on the spontaneous emission rate $\Gamma_{g}^{ns}$
into the guided modes for an atom initially prepared in the state
$\left|ns_{\nicefrac{1}{2}},m_{j}=\pm\frac{1}{2}\right\rangle $ for
$n=5$ to $10$. The higher the value of $n$, the more $\Gamma_{g}^{ns}$
is peaked as a function of $r/a$ around $1$. Moreover, the plots
get closer and closer as $n$ increases : the curves $n=9,10$ cannot
be distinguished and for the sake of clarity, the curve $n=8$ has
not been plotted.

\begin{figure*}
\begin{centering}
\includegraphics[width=14cm]{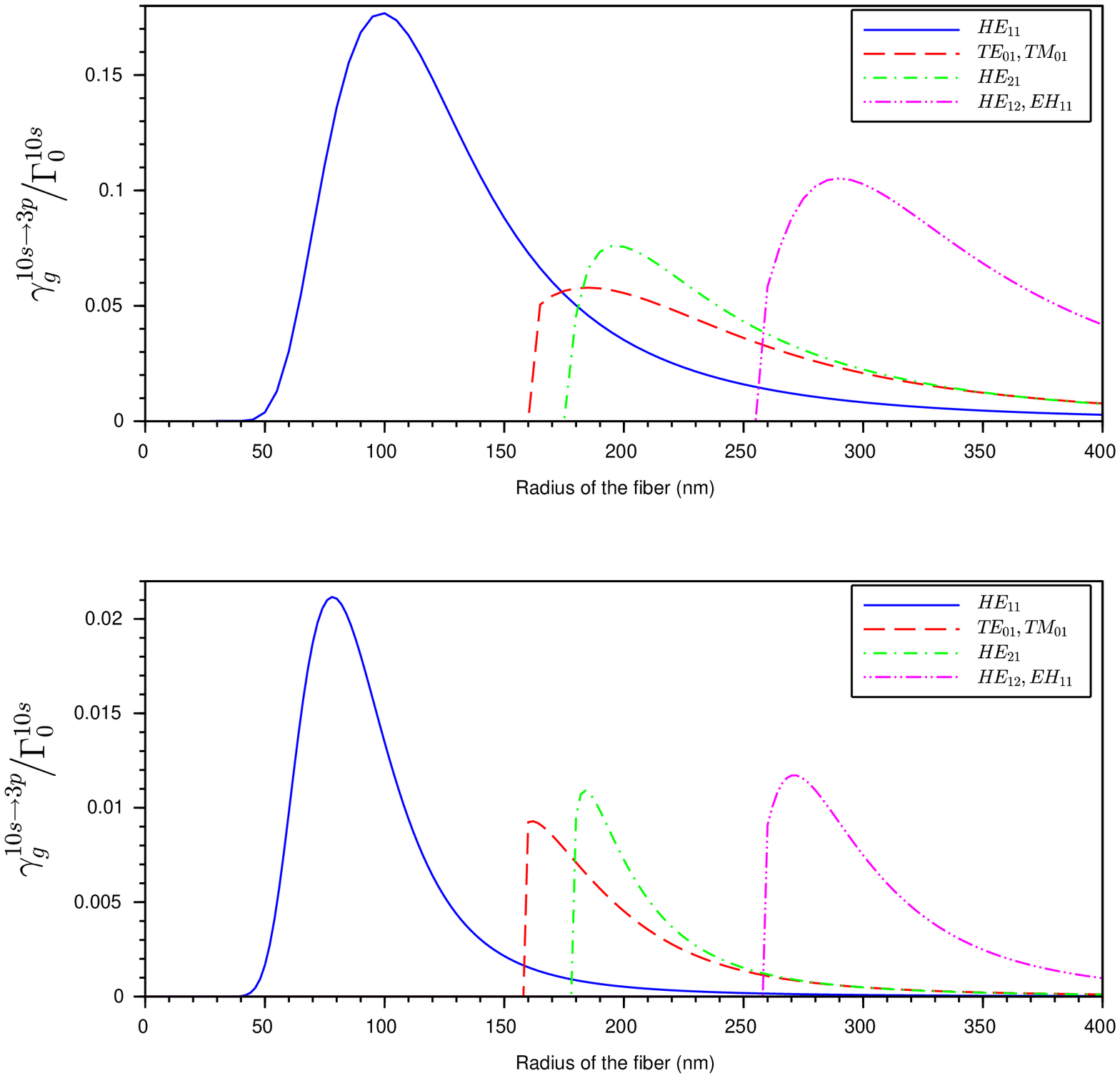}
\par\end{centering}
\caption{Partial spontaneous emission rates of a sodium atom along the specific
transition $10s\rightarrow3p$ into different guided modes of the
nanofiber, as functions of the fibre radius, $a$ : (top) the atom
is on the fibre surface ; (bottom) the atom is located at a distance
$150$ nm from the fibre surface. The rates are presented relative
to the spontaneous emission rate in vacuum, $\Gamma_{0}^{10s}$.}

\label{FigModGuidRayonFibrePartiel}
\end{figure*}

\begin{figure*}
\begin{centering}
\includegraphics[width=14cm]{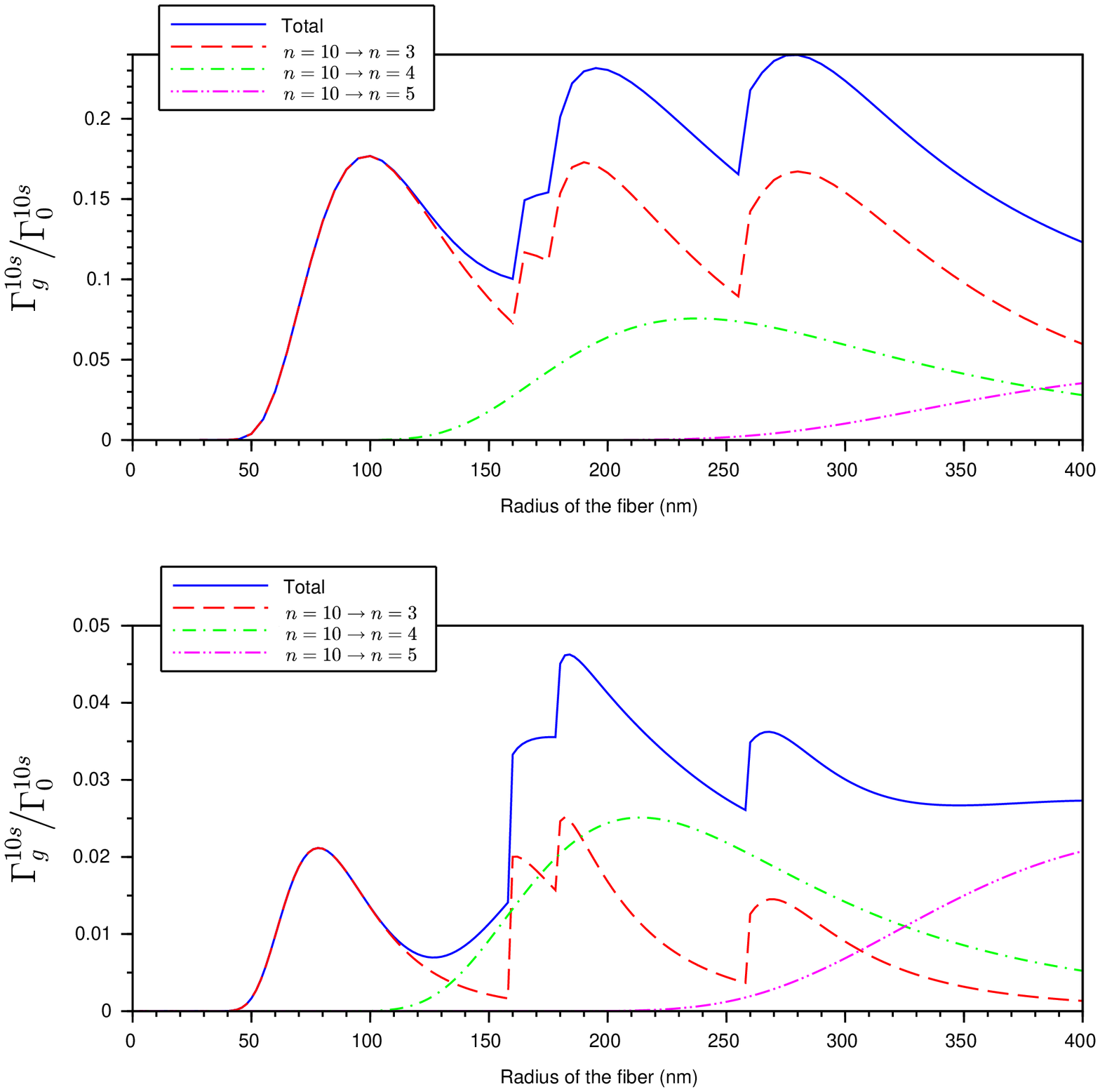}
\par\end{centering}
\caption{Partial, $\gamma_{g}^{10s\rightarrow np}$, and total, $\Gamma_{g}^{10s}$,
spontaneous emission rates of a sodium atom initially prepared in
$\left|10s_{\nicefrac{1}{2}},m_{j}=\pm\frac{1}{2}\right\rangle $
into the guided modes of a nanofibre. The rates are plotted as a function
of the fibre radius, $a$ : (top) the atom is on the fibre surface
; (bottom) the atom is located at a distance $150$ nm from the fibre
surface. The rates are presented relative to the spontaneous emission
rate in vacuum, $\Gamma_{0}^{10s}$.}

\label{FigModGuidRayonFibreTotal}
\end{figure*}

Finally, figures \ref{FigModGuidRayonFibrePartiel} and \ref{FigModGuidRayonFibreTotal}
illustrate the influence of the fibre radius, $a$, on the spontaneous
emission rate from the state $\left|ns_{\nicefrac{1}{2}},m_{j}=\pm\frac{1}{2}\right\rangle $
into the guided modes. More precisely, figure \ref{FigModGuidRayonFibrePartiel}
displays the partial spontaneous emission rates along the specific
transition $10s\rightarrow3p$ (Note that $n=3$ corresponds to the
ground state of the sodium atom) into different guided modes $HE_{mn}$,
$EH_{mn}$, $TE_{mn}$ and $TM_{mn}$. Two cases are considered :
(\emph{i}) the atom is located on the fibre surface, i.e. at a distance
$r=a$ from the $z$-axis, and (\emph{ii}) the atom is placed at a
fixed distance of $150$ nm from the fibre surface, i.e. at a distance
$r=a+150\text{ nm}$ from the $z$-axis. As expected, case (\emph{ii})
gives rise to much weaker relative rates than case (\emph{i}), since
the atom is further away from the fibre and therefore the guided modes
are strongly attenuated. Moreover, as $a$ increases, the cutoff frequencies
of higher modes become smaller : when the cutoff frequency of one
mode passes below the frequency of the transition $10s\rightarrow3p$,
this mode starts to contribute to the spontaneous emission rate. The
peaked structure observed on the different plots results from the
peaked shape of the mode intensity profile itself with respect to
$a$.

Figure \ref{FigModGuidRayonFibreTotal} displays the partial spontaneous
emission rates $\gamma_{g}^{10s\rightarrow np}$ into the guided modes
along the respective transitions $10s\rightarrow np$ as well as the
total rate $\Gamma_{g}^{10s}=\sum_{3\leq n\leq10}\gamma_{g}^{10s\rightarrow np}$
as functions of the fibre radius, $a$, in the same two cases (\emph{i},
\emph{ii}) as above. One observes that, due to the range chosen for
$a$, only the transitions $10s\rightarrow np$ for $n=3,4,5$ give
relevant contributions to the total rate. It also appears that only
the transition $10s\rightarrow3p$ substantially couples to higher-order
guided modes, while the other transitions couple only to the fundamental
guided mode $HE_{11}$. On the range chosen for $a$, the peak structure
observed for the total emission rate is therefore mainly due to the
partial rate $\gamma_{g}^{10s\rightarrow3p}$, while the other transitions
smoothly modify the value of $\Gamma_{g}^{10s}$. Note that the intensity
profiles of the guided modes relative to the different transition
frequencies are expected to coincide up to a rescaling of the $a$-axis
: this scaling factor is given by the ratio of the frequencies. The
positions of the peaks of the different partial rates $\gamma_{g}^{10s\rightarrow np}$
should therefore also coincide up to a simple scaling. The heights
of the peaks, however, are expected to be different since, for instance,
the dipole matrix element is not the same for the different transitions.

\subsection{Radiative modes}

\begin{figure*}
\begin{centering}
\includegraphics[width=14cm]{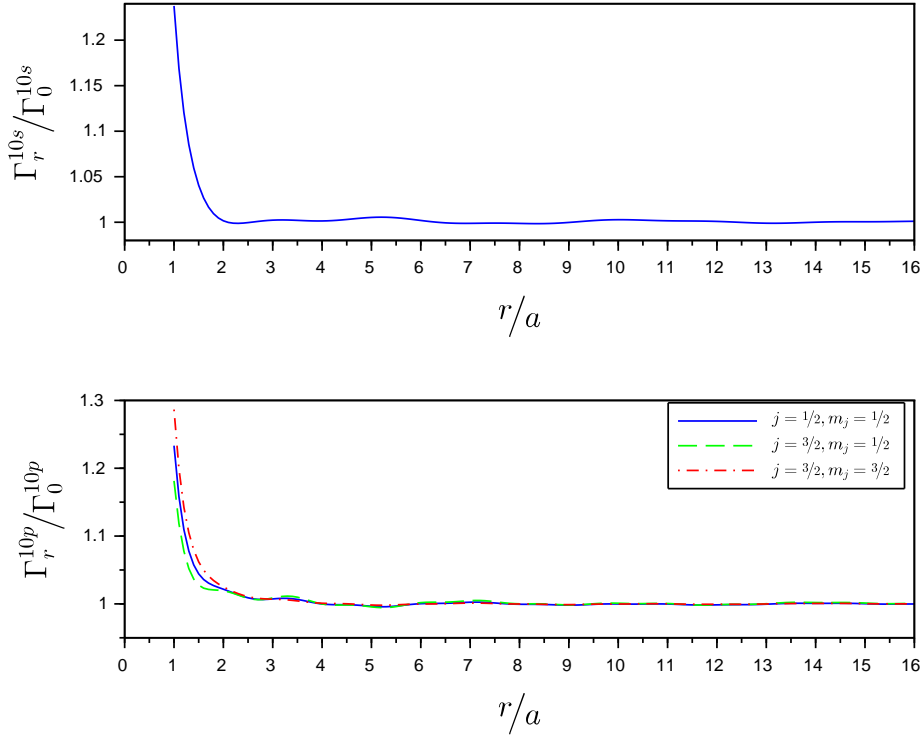}
\par\end{centering}
\caption{Total spontaneous emission rate of a sodium atom into the radiative
modes of a nanofibre of radius $a=100$ nm. The rate is plotted as
a function of the distance $r$ of the atom to the fibre axis : (top)
atom initially prepared in the state $\left|10s_{\nicefrac{1}{2}},m_{j}=\pm\frac{1}{2}\right\rangle $,
(bottom) atom initially prepared in the states $\left|10p_{\nicefrac{1}{2}},m_{j}=\pm\frac{1}{2}\right\rangle $,
$\left|10p_{\nicefrac{3}{2}},m_{j}=\pm\frac{1}{2}\right\rangle $
and $\left|10p_{\nicefrac{3}{2}},m_{j}=\pm\frac{3}{2}\right\rangle $.
The rates $\Gamma_{r}^{10s},\Gamma_{r}^{10p}$ are presented relative
to the spontaneous emission rates $\Gamma_{0}^{10s}$, $\Gamma_{0}^{10p}$
in vacuum and the distance $r$ is expressed in units of the fibre
radius, $a$.}
\label{FIGRad10s10p}
\end{figure*}

We now turn to the contribution of the radiative modes to the total
spontaneous emission rates. Figure \ref{FIGRad10s10p} displays the
spontaneous emission rates $\Gamma_{r}^{10s}$ and $\Gamma_{r}^{10p}$
of an atom initially prepared in the states $\left|10s_{\nicefrac{1}{2}},m_{j}\right\rangle $,
$\left|10p_{\nicefrac{1}{2},\nicefrac{3}{2}},m_{j}\right\rangle $,
respectively into the \emph{radiative} modes as a function of the
distance $r$ to the fibre axis (see figure \ref{System}). Note that
the rates are renormalized by the spontaneous emission rates in vacuum
$\Gamma_{0}^{10s}$, resp. $\Gamma_{0}^{10p}$, and $r$ is expressed
in units of the fibre radius $a=100$ nm. As expected, in both cases,
the influence of the fibre vanishes as $r$ increases, i.e. $\Gamma_{r}^{10s},\Gamma_{r}^{10p}\rightarrow1$
for $r\rightarrow+\infty$. The maximal value is observed for $r=a$,
i.e. when the atom is on the fibre surface. More precisely, we have
$\Gamma_{r}^{10s}\approx1.24\times\Gamma_{0}^{10s}$ for an atom initially
prepared in $\left|10s_{\nicefrac{1}{2}},m_{j}=\pm\frac{1}{2}\right\rangle $
and $\Gamma_{r}^{10p}\approx\left(1.19,1.23,1.29\right)\times\Gamma_{0}^{10p}$
for an atom initially prepared in $\left(\left|10p_{\nicefrac{3}{2}},m_{j}=\pm\frac{1}{2}\right\rangle ,\left|10p_{\nicefrac{1}{2}},m_{j}=\pm\frac{1}{2}\right\rangle ,\left|10p_{\nicefrac{3}{2}},m_{j}=\pm\frac{3}{2}\right\rangle \right)$.
For an atom at $r\approx1.5\times a$, , i.e., at a distance from
the fibre surface, we obtain the spontaneous rate $\Gamma_{r}^{10s}\approx1.041\times\Gamma_{0}^{10s}$
for an atom initially prepared in $\left|10s_{\nicefrac{1}{2}},m_{j}=\pm\frac{1}{2}\right\rangle $
and $\Gamma_{r}^{10p}\approx\left(1.028,1.044,1.062\right)\times\Gamma_{0}^{10p}$
for an atom initially prepared in $\left(\left|10p_{\nicefrac{3}{2}},m_{j}=\pm\frac{1}{2}\right\rangle ,\left|10p_{\nicefrac{1}{2}},m_{j}=\pm\frac{1}{2}\right\rangle ,\left|10p_{\nicefrac{3}{2}},m_{j}=\pm\frac{3}{2}\right\rangle \right)$.
This allows us to compute the proportion of light which is emitted
into the guided and radiative modes. For instance, for an atom initially
prepared in the state $\left|10s_{\nicefrac{1}{2}},m_{j}=\pm\frac{1}{2}\right\rangle $,
$\Gamma_{g}/\left(\Gamma_{g}+\Gamma_{r}\right)\approx13\%$ when the
atom is located on the fibre surface $\left(r=a\right)$, and $\Gamma_{g}/\left(\Gamma_{g}+\Gamma_{r}\right)\approx6\%$
when the atom is located at $50$ nm from the fibre surface $\left(r=1.5\times a\right)$.
Since light is mostly spontaneously emitted into the radiative modes,
it seems quite challenging to efficiently interface a Rydberg atom
with a guided mode of the nanofibre and, thence, to build a valuable
quantum network. The use of atomic ensembles might alleviate this
concern, since, as already demonstrated in free-space, their spontaneous
emission could be made highly directional and their coupling strength
is enhanced \citep{PM09}. These issues and the perspectives they
offer will be addressed in a future work. 

\begin{figure*}
\begin{centering}
\includegraphics[width=14cm]{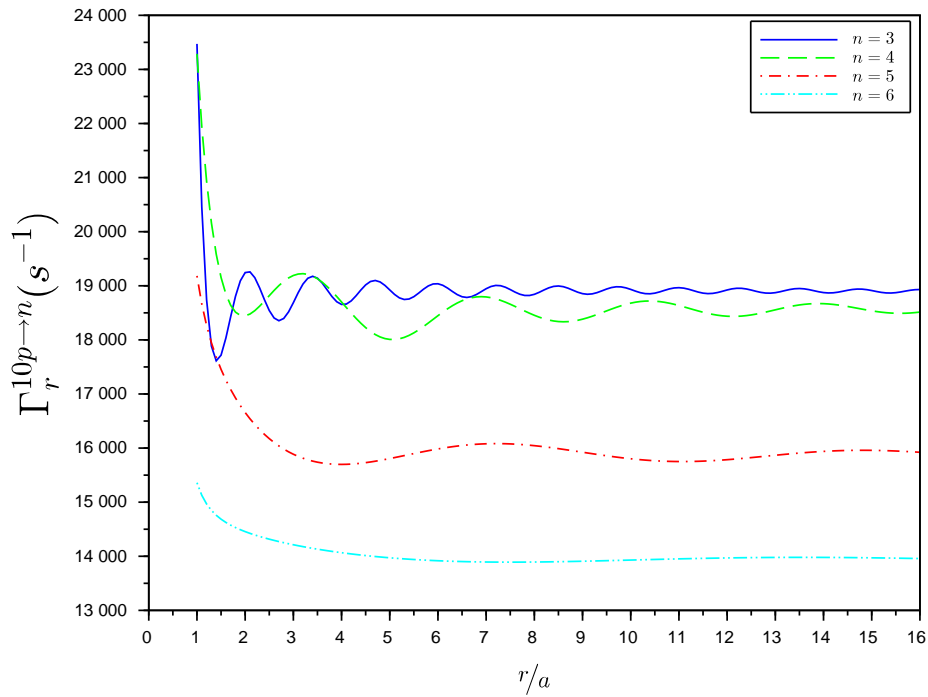}
\par\end{centering}
\caption{Spontaneous emission rate of a sodium atom initially prepared in $\left|10p_{\nicefrac{3}{2}},m_{j}=\pm\frac{1}{2}\right\rangle $
into the radiative modes of a nanofibre of radius $a=100$ nm : contributions
of the different transitions $\left|10p_{\nicefrac{3}{2}},m_{j}=\pm\frac{1}{2}\right\rangle \rightarrow\left|ns_{\nicefrac{1}{2}},m_{j}=\pm\frac{1}{2}\right\rangle ,\left|nd_{\nicefrac{5}{2}},m_{j}=\pm\frac{1}{2},\pm\frac{3}{2}\right\rangle $,
for $n=3,\cdots,6$, . The rate is plotted as a function of the distance
$r$ of the atom to the fibre axis. The distance $r$ is expressed
in units of the fibre radius, $a$. \label{FIGRadTrans}}
\end{figure*}

Finally, in figure \ref{FIGRad10s10p}, one observes a damped semi-oscillatory
behaviour for $\Gamma_{r}^{10s}$ and $\Gamma_{r}^{10p}$ as functions
of $r$, and for $\Gamma_{r}^{10p}$ the oscillations of the different
contributions $j=1/2,3/2$ are not in phase. These features result
from the behaviour of the different transition components $\gamma_{nl\rightarrow n'l'}$
shown in figure \ref{FIGRadTrans} for $nl=np_{\nicefrac{3}{2}},m_{j}=\pm\frac{1}{2}$,
which is itself due to the oscillatory behaviour of the radiative
field. For a transition of frequency $\omega$, the frequency of oscillation
with $r$ is approximately given by $2\omega/c$. 

\section{Conclusion \label{SECConclusion}}

We have investigated the influence of an optical nanofibre on the
spontaneous emission rate of a sodium atom prepared in a Rydberg state.
The respective contributions of the guided and radiative modes to
the total rate were numerically determined, for different principal
quantum numbers and different symmetries, and their remarkable features
were physically discussed.

Though the radiative modes' contribution is dominant, a small fraction
of the spontaneously emitted light is transferred into the guided
mode of the nanofibre. This effect might be enhanced by resorting
to atomic ensembles which could offer stronger and more directional
collective coupling. Using thicker fibres, with more than one guided
mode, may also yield for a higher ratio of spontaneous emission into
the guided modes. This potentially paves the way towards the implementation
of a quantum network based on Rydberg atomic ensembles linked by nanofibres,
which will be further addressed in a future work. 

\ack{}{This research was supported by the Centre National de la Recherche
Scientifique (CNRS) via the grant ``PICS QuaNet''. SNC acknowledges
support from OIST Graduate University. The authors thank Antoine Browaeys,
Tridib Ray and Fam Le Kien for fruitful discussions. }

\appendix

\section{Guided modes \label{APPModesGuides}}

A guided mode is characterized by a set $\mu\equiv\left(\omega,\beta,m,f=\pm,p=\pm\right)$.
$\beta$ is the projection of the wavevector onto the axis of the
nanofibre whose value is determined by the eigenvalue equation
\begin{eqnarray}
\left(\frac{n_{1}^{2}J_{m}'(\kappa a)}{a\kappa J_{m}(\kappa a)}+\frac{n_{2}^{2}K_{m}'(\gamma a)}{a\gamma K_{m}(\kappa a)}\right)\left(\frac{J_{m}'(\kappa a)}{a\kappa J_{m}(\kappa a)}+\frac{K_{m}'(\gamma a)}{a\gamma K_{m}(\gamma a)}\right)\nonumber \\
=\left(\frac{mc\beta}{\omega}\right)^{2}\left(\frac{1}{\left(\gamma a\right)^{2}}+\frac{1}{\left(\kappa a\right)^{2}}\right)^{2}.\label{EQCaracteristique}
\end{eqnarray}
Here we introduced $\kappa\equiv\sqrt{n_{1}^{2}k^{2}-\beta^{2}}$,
$\gamma\equiv\sqrt{\beta^{2}-n_{2}^{2}k^{2}}$ and $k\equiv\frac{\omega}{c}$.
$a$ is the radius of the fibre, $n_{1}$ is the core index, $n_{2}\approx1$
is the index of the surrounding vacuum. $J_{m}$ and $K_{m}$ denote
the Bessel functions of the first kind and the modified Bessel functions
of the second kind, respectively. Note that, when the monomode conditions
are fulfilled, only the hybrid modes $HE_{11}$ with $m=1$ exist,
and are fully characterized by $\mu\equiv\left(\omega,f,p\right)$.

The polarization vectors of the guided mode $\left(\mu\right)$ for
$r<a$ are given by 
\begin{eqnarray*}
e_{r}^{\left(\mu\right)} & = & \frac{\beta C}{2\textrm{i}\kappa}\frac{K_{m}(\gamma a)}{J_{m}(\kappa a)}\Big(J_{m-1}(\kappa r)(1-ms)-J_{m+1}(\kappa r)(1+ms)\Big)\\
e_{\varphi}^{\left(\mu\right)} & = & \frac{p\beta C}{2\kappa}\frac{K_{m}(\gamma a)}{J_{m}(\kappa a)}\Big(J_{m-1}(\kappa r)(1-ms)+J_{m+1}(\kappa r)(1+ms)\Big),\\
e_{z}^{\left(\mu\right)} & = & C\frac{K_{m}(\gamma a)}{J_{m}(\kappa a)}J_{m}(\kappa r),
\end{eqnarray*}
while, for $r>a$, they are
\begin{eqnarray*}
e_{r}^{\left(\mu\right)} & = & \frac{\beta C}{2\textrm{i}\gamma}\Big(K_{m-1}(\gamma r)(1-ms)+K_{m+1}(\gamma r)(1+ms)\Big),\\
e_{\varphi}^{\left(\mu\right)} & = & \frac{\beta pC}{2\gamma}\Big(K_{m-1}(\gamma r)(1-ms)-K_{m+1}(\gamma r)(1+ms)\Big),\\
e_{z}^{\left(\mu\right)} & = & CK_{m}(\gamma r),
\end{eqnarray*}
where
\[
s=\frac{\frac{1}{\gamma^{2}a^{2}}+\frac{1}{\kappa^{2}a^{2}}}{\frac{J_{m}'(\kappa a)}{\kappa aJ_{m}(\kappa a)}+\frac{K_{m}'(\gamma a)}{\gamma aK_{m}(\gamma a)}}.
\]
Using the normalization condition 

\[
\int_{0}^{2\pi}\mathrm{d}\varphi\int_{0}^{+\infty}n\left(r\right)^{2}\left|{\bf e}^{\left(\mu\right)}\right|^{2}~r\mathrm{d}r=1,
\]
we deduce that 

\begin{eqnarray*}
|C| & = & \frac{2}{a\beta K_{m}\left(\gamma a\right)\sqrt{2\pi(n_{1}^{2}A_{1}+n_{2}^{2}A_{2})}}
\end{eqnarray*}
with the abbreviations 
\begin{eqnarray*}
A_{1} & = & \left(\frac{1}{\kappa J_{m}\left(\kappa a\right)}\right)^{2}\times\left(\left(1-ms\right)^{2}\left(J_{m-1}^{2}\left(\gamma a\right)-J_{m}\left(\gamma a\right)J_{m-2}\left(\gamma a\right)\right)\right.\\
 &  & +(1+ms)^{2}\left(J_{m+1}^{2}\left(\gamma a\right)-J_{m}\left(\gamma a\right)J_{m+2}\left(\gamma a\right)\right)\\
 &  & \left.+2\frac{\kappa^{2}}{\beta^{2}}\left(J_{m}^{2}\left(\gamma a\right)-J_{m-1}\left(\gamma a\right)J_{m+1}\left(\gamma a\right)\right)\right),\\
A_{2} & = & \left(\frac{1}{\gamma K_{m}\left(\gamma a\right)}\right)^{2}\times\left(\left(1-ms\right)^{2}\left(-K_{m-1}^{2}\left(\gamma a\right)+K_{m}\left(\gamma a\right)K_{m-2}\left(\gamma a\right)\right)\right.\\
 &  & +\left(1+ms\right)^{2}\left(-K_{m+1}^{2}\left(\gamma a\right)+K_{m}\left(\gamma a\right)K_{m+2}\left(\gamma a\right)\right)\\
 &  & \left.+2\left(\frac{\gamma}{\beta}\right)^{2}\left(-K_{m}^{2}\left(\gamma a\right)+K_{m-1}\left(\gamma a\right)K_{m+1}\left(\gamma a\right)\right)\right).
\end{eqnarray*}

\section{Radiative modes\label{APPModesRadiatifs}}

A radiative mode is characterized by a set $\nu\equiv\left(\omega,\beta,m,p=\pm\right)$,
where $m$ is the order of the mode, and the meaning of $p$ will
be explained below. 

Defining the quantities $\kappa\equiv\sqrt{n_{1}^{2}k^{2}-\beta^{2}}$,
$\sigma\equiv\sqrt{n_{2}^{2}k^{2}-\beta^{2}}$ and $k\equiv\omega/c$,
one can write the polarization vectors of the radiative mode $\left(\nu\right)$
for $r<a$ :

\begin{eqnarray*}
e_{r}^{\left(\nu\right)} & = & \frac{1}{\textrm{i}\kappa}\Big(\beta AJ_{m}'(\kappa r)+\mathrm{i}B\frac{\omega m}{r\kappa}J_{m}(\kappa r)\Big),\\
e_{\varphi}^{\left(\nu\right)} & = & \frac{1}{\textrm{i}\kappa}\Big(\mathrm{i}A\frac{m\beta}{\kappa r}J_{m}(\kappa r)-\omega BJ_{m}'(\kappa r)\Big),\\
e_{z}^{\left(\nu\right)} & = & AJ_{m}(\kappa r),
\end{eqnarray*}
while for $r>a$ :

\begin{eqnarray*}
e_{r}^{\left(\nu\right)} & = & \frac{1}{\textrm{i}\sigma}\left[\beta\left(CJ_{m}'(\sigma r)+EY_{m}'(\sigma r)\right)+\frac{\mathrm{i}m\omega}{\sigma r}\left(DJ_{m}\left(\sigma r\right)+FY_{m}\left(\sigma r\right)\right)\right],\\
e_{\varphi}^{\left(\nu\right)} & = & \frac{1}{\textrm{i}\sigma}\left[\mathrm{i}\frac{\nu\beta}{\sigma r}\left(CJ_{m}\left(\sigma r\right)+EY_{m}\left(\sigma r\right)\right)-\omega\left(DJ_{m}'(\sigma r)+FY_{m}'(\sigma r)\right)\right],\\
e_{z}^{\left(\nu\right)} & = & CJ_{m}\left(\sigma r\right)+EY_{m}\left(\sigma r\right),
\end{eqnarray*}
where $Y_{m}$ denote Bessel functions of the second kind. The coefficients
$C$, $D$, $E$ and $F$ are related to $A$ and $B$ as follows
\begin{eqnarray*}
C & = & -\frac{\pi a\sigma^{2}}{2\epsilon_{0}n_{2}^{2}}\left(AL_{2}+\textrm{i}BV_{2}\right),\\
D & = & \textrm{i}\frac{\pi a\sigma^{2}}{2}\left(A\mu_{0}V_{2}+\textrm{i}BM_{2}\right),\\
E & = & \frac{\pi a\sigma^{2}}{2\epsilon_{0}n_{2}^{2}}\left(AL_{1}+\textrm{i}BV_{1}\right),\\
F & = & -\textrm{i}\frac{\pi a\sigma^{2}}{2}\left(A\mu_{0}V_{1}+\textrm{i}BM_{1}\right),
\end{eqnarray*}
with
\begin{eqnarray*}
V_{1} & = & \frac{m\beta}{a\omega\mu_{0}\kappa^{2}\sigma^{2}}\left(n_{2}^{2}-n_{1}^{2}\right)J_{m}\left(\sigma a\right)J_{m}\left(\kappa a\right),\\
V_{2} & = & \frac{m\beta}{a\omega\mu_{0}\kappa^{2}\sigma^{2}}\left(n_{2}^{2}-n_{1}^{2}\right)Y_{m}\left(\sigma a\right)J_{m}\left(\kappa a\right),\\
M_{1} & = & \frac{1}{\kappa}J_{m}\left(\sigma a\right)J_{m}'\left(\kappa a\right)-\frac{1}{\sigma}J_{m}'\left(\sigma a\right)J_{m}\left(\kappa a\right),\\
M_{2} & = & \frac{1}{\kappa}Y_{m}\left(\sigma a\right)J_{m}'\left(\kappa a\right)-\frac{1}{\sigma}Y_{m}'\left(\sigma a\right)J_{m}\left(\kappa a\right),\\
L_{1} & = & \frac{\epsilon_{0}n_{1}^{2}}{\kappa}J_{m}\left(\sigma a\right)J_{m}'\left(\kappa a\right)-\frac{\epsilon_{0}n_{2}^{2}}{\sigma}J_{m}'\left(\sigma a\right)J_{m}\left(\kappa a\right),\\
L_{2} & = & \frac{\epsilon_{0}n_{1}^{2}}{\kappa}Y_{m}\left(\sigma a\right)J_{m}'\left(\kappa a\right)-\frac{\epsilon_{0}n_{2}^{2}}{\sigma}Y_{m}'\left(\sigma a\right)J_{m}\left(\kappa a\right).
\end{eqnarray*}
In the single-mode approximation, a guided mode is completely specified
by the frequency $\omega$, the direction of propagation $f=\pm1$
and the polarization $p=\pm1$. By contrast, at first glance, this
is not the case for the radiative modes any longer. Once $\beta$,
$\omega$ and $m$ are fixed, we are left with 2 constants $A$ and
$B$, and a normalization condition will only determine one constant.
We must therefore separate these into two modes. For instance, we
can just set $A=0$ for one mode and $B=0$ for the other one. We
want, however, the two modes to be orthogonal to each other. An alternative
method consists in setting $B=p\mathrm{i}\eta A$ with the parameter
$p=\pm1$, then imposing an orthogonality condition between ${\bf e}^{\left(p=+1\right)}$
and ${\bf e}^{\left(p=-1\right)}$. Explicitly, this condition is
written~: 
\begin{eqnarray*}
\int_{0}^{2\pi}d\varphi\int_{0}^{\infty}n\left(r\right)^{2}\left[{\bf e}^{\left(\nu\right)}\cdot{\bf e}^{\left(\nu'\right)}\right]rdr & = & \delta_{pp'}\delta_{mm'}\delta\left(\omega-\omega'\right)\delta\left(\beta-\beta'\right).
\end{eqnarray*}
If we consider the vacuum surrounding with the index $n_{2}=1$, this
leads to~:

\begin{eqnarray*}
\eta & = & \sqrt{\frac{L_{1}^{2}+L_{2}^{2}+\epsilon_{0}\mu_{0}\left(V_{1}^{2}+V_{2}^{2}\right)}{V_{1}^{2}+V_{2}^{2}+\frac{\epsilon_{0}}{\mu_{0}}\left(M_{1}^{2}+M_{2}^{2}\right)}},\\
1 & = & \frac{2\pi\omega}{\sigma^{2}}\left[\left(\left|C\right|^{2}+\left|E\right|^{2}\right)+c^{2}\left(\left|D\right|^{2}+\left|F\right|^{2}\right)\right].
\end{eqnarray*}
The second normalization equation allows us to calculate the form
of $\left|A\right|$

\[
\left|A\right|=\frac{1}{\sigma a}\left[\frac{2}{\omega\pi^{3}}\sum_{j=1,2}\left(\frac{1}{\epsilon_{0}}\left|L_{j}-p\eta V_{j}\right|^{2}+\left|\mu_{0}V_{j}-p\eta M_{j}\right|^{2}\right)\right]^{-\frac{1}{2}}.
\]
This shows that $\nu=\left(\omega,\beta,m,p\right)$ completely determines
a radiative mode. 

\section{Spontaneous emission of an atom in the presence of a nanofibre\label{APPRates}}

With the definitions $\sigma_{MN}\equiv\left|M\right\rangle \left\langle N\right|$,
$\omega_{MN}\equiv\left(E_{M}-E_{N}\right)/\hbar$, $k\equiv\omega/c$,
the Hamiltonian of the full system consisting of the atom and the
electric field takes the form $H=H_{at}+H_{f}+H_{int}$ with

\begin{eqnarray*}
H_{f} & = & \sum_{\mu}\hbar\omega a_{\mu}^{\dagger}a_{\mu}+\sum_{\nu}\hbar\omega a_{\nu}^{\dagger}a_{\nu},\\
H_{at} & = & \sum_{m}\hbar\omega_{M}\sigma_{MM},\\
H_{int} & = & -\vec{D}\cdot\vec{E},
\end{eqnarray*}
where $\vec{D}=\sum_{M,N}\vec{d}_{MN}\sigma_{MN}$ and $\vec{E}\left(\vec{r}\right)=\vec{E}_{g}\left(\vec{r}\right)+\vec{E}_{r}\left(\vec{r}\right)$
are the atomic dipole operator and the total\textbf{ }electric field
operator, respectively. Switching to the interaction picture relative
to $H_{0}\equiv H_{at}+H_{f}$, and resorting to the rotating wave
approximation (RWA) we get the interaction Hamiltonian (note that
the states $\left|M\right\rangle $ are ordered by increasing energies)
\begin{eqnarray*}
\tilde{H}_{int}\left(t\right) & \approx & \mathrm{i}\hbar\sum_{N<M}\sum_{\mu}\;G_{\mu MN}\sigma_{MN}a_{\mu}e^{-\mathrm{i}\left(\omega-\omega_{MN}\right)t}+\mathrm{h.c.}\\
 &  & +\mathrm{i}\hbar\sum_{N<M}\sum_{\nu}\;G_{\nu MN}\sigma_{MN}a_{\nu}e^{-\mathrm{i}\left(\omega-\omega_{MN}\right)t}+\mathrm{h.c.}
\end{eqnarray*}
where we introduced
\begin{eqnarray*}
G_{\mu MN} & \equiv & -\sqrt{\frac{\omega\beta'}{4\pi\epsilon_{0}\hbar}}\left(\vec{d}_{MN}\cdot\vec{e}^{\left(\mu\right)}\right)e^{\mathrm{i}\left(f\beta z+p\varphi\right)},\\
G_{\nu MN} & \equiv & -\sqrt{\frac{\omega}{4\pi\epsilon_{0}\hbar}}\left(\vec{d}_{MN}\cdot\vec{e}^{\left(\nu\right)}\right)e^{\mathrm{i}\left(\beta z+m\varphi\right)},
\end{eqnarray*}
For simplicity, from now on we shall use $\lambda$ to denote either
guided modes, i.e. $\lambda=\left(\omega,f=\pm,p=\pm\right)$, or
radiative modes, i.e. $\lambda=\left(\omega,\beta,m,p\right)$, and
use $\sum_{\lambda}$ to represent the sum, either discrete or continuous,
of these modes, whence
\begin{eqnarray}
\tilde{H}_{int}\left(t\right) & \approx & \mathrm{i}\hbar\sum_{N<M}\sum_{\lambda}\;G_{\lambda MN}\sigma_{MN}a_{\lambda}e^{-\mathrm{i}\left(\omega_{\lambda}-\omega_{MN}\right)t}\nonumber \\
 &  & -\mathrm{i}\hbar\sum_{N<M}\sum_{\lambda}\;G_{\lambda MN}^{*}a_{\lambda}^{\dagger}\sigma_{NM}e^{+\mathrm{i}\left(\omega_{\lambda}-\omega_{MN}\right)t}\label{EQHAM-1}
\end{eqnarray}

From equation (\ref{EQHAM-1}), we get the Heisenberg equations for
the field and atomic operators, $a_{\lambda}$ and $\sigma_{PQ}$ 

\begin{eqnarray}
\partial_{t}a_{\lambda} & = & -\sum_{N<M}\;G_{\lambda MN}^{*}\sigma_{NM}e^{+\mathrm{i}\left(\omega_{\lambda}-\omega_{MN}\right)t},\label{EQHa-1}\\
\partial_{t}\sigma_{PQ} & = & \sum_{M<Q}\sum_{\lambda}\;G_{\lambda QM}\sigma_{PM}\left(t\right)a_{\lambda}\left(t\right)e^{-\mathrm{i}\left(\omega_{\lambda}-\omega_{QM}\right)t}\nonumber \\
 &  & -\sum_{P<M}\sum_{\lambda}\;G_{\lambda MP}\sigma_{MQ}\left(t\right)a_{\lambda}\left(t\right)e^{-\mathrm{i}\left(\omega_{\lambda}-\omega_{MP}\right)t}\nonumber \\
 &  & -\sum_{Q<M}\sum_{\lambda}\;G_{\lambda MQ}^{*}a_{\lambda}^{\dagger}\left(t\right)\sigma_{PM}\left(t\right)e^{+\mathrm{i}\left(\omega_{\lambda}-\omega_{MQ}\right)t}\nonumber \\
 &  & +\sum_{M<P}\sum_{\lambda}\;G_{\lambda PM}^{*}a_{\lambda}^{\dagger}\left(t\right)\sigma_{MQ}\left(t\right)e^{+\mathrm{i}\left(\omega_{\lambda}-\omega_{PM}\right)t}.\label{EQHs-1}
\end{eqnarray}
We can eliminate the field degree of freedom by inserting the formal
solution of equation (\ref{EQHa-1}) 
\begin{eqnarray*}
a_{\lambda}\left(t\right) & = & a_{\lambda}\left(t_{0}\right)-\sum_{N<M}\;G_{\lambda MN}^{*}\int_{t_{0}}^{t}\sigma_{NM}\left(s\right)e^{+\mathrm{i}\left(\omega_{\lambda}-\omega_{MN}\right)s}
\end{eqnarray*}
into equation (\ref{EQHs-1}). Then performing Markov approximation
\citep{WM08} and using 
\begin{eqnarray*}
\int_{0}^{\infty}dse^{\pm\mathrm{i}\Delta s} & = & \pi\delta\left(\Delta\right)\pm\mathrm{i}\mathcal{P}\left(\frac{1}{\Delta}\right)
\end{eqnarray*}
we get

\begin{eqnarray*}
\partial_{t}\sigma_{PQ} & \approx & -\left\{ \Gamma_{PQ}+\mathrm{i}\Delta_{PQ}\right\} \sigma_{PQ}\left(t\right)+\delta_{PQ}\sum_{P<M}\gamma_{MP}\sigma_{MM}\left(t\right)+\xi_{PQ}\left(t\right)
\end{eqnarray*}
where we introduced the different decay rates and energy shifts due
to spontaneous emission into the modes of the fibre,
\begin{eqnarray*}
\Gamma_{PQ} & \equiv & \frac{1}{2}\left(\Gamma_{P}+\Gamma_{Q}\right)\\
\Delta_{PQ} & \equiv & \Delta_{P}-\Delta_{Q}\\
\Gamma_{P} & \equiv & \sum_{M<P}\gamma_{PM}\\
\gamma_{PQ} & \equiv & 2\pi\sum_{\lambda}\left|G_{\lambda PQ}\right|^{2}\delta\left(\omega_{\lambda}-\omega_{PQ}\right)\\
\Delta_{P} & \equiv & \mathcal{P}\left(\sum_{M<P}\sum_{\lambda}\frac{\left|G_{\lambda PM}\right|^{2}}{\omega_{\lambda}-\omega_{PM}}-\frac{\left|G_{\lambda PM}^{\left(0\right)}\right|^{2}}{\omega_{\lambda}-\omega_{PM}}\right)+\Delta_{P}^{\left(0\right)}
\end{eqnarray*}
and the associated Langevin forces
\begin{eqnarray}
\xi_{PQ}\left(t\right) & \equiv & \sum_{M<Q}\sum_{\lambda}\;G_{\lambda QM}\sigma_{PM}\left(t\right)a_{\lambda}\left(t_{0}\right)e^{-\mathrm{i}\left(\omega_{\lambda}-\omega_{QM}\right)t}\nonumber \\
 &  & -\sum_{P<M}\sum_{\lambda}\;G_{\lambda MP}\sigma_{MQ}\left(t\right)a_{\lambda}\left(t_{0}\right)e^{-\mathrm{i}\left(\omega_{\lambda}-\omega_{MP}\right)t}\nonumber \\
 &  & -\sum_{Q<M}\sum_{\lambda}\;G_{\lambda MQ}^{*}a_{\lambda}^{\dagger}\left(t_{0}\right)\sigma_{PM}\left(t\right)e^{+\mathrm{i}\left(\omega_{\lambda}-\omega_{MQ}\right)t}\nonumber \\
 &  & +\sum_{M<P}\sum_{\lambda}\;G_{\lambda PM}^{*}a_{\lambda}^{\dagger}\left(t_{0}\right)\sigma_{MQ}\left(t\right)e^{+\mathrm{i}\left(\omega_{\lambda}-\omega_{PM}\right)t}.\label{EQxi-1}
\end{eqnarray}
Note that due to the normal operator ordering in equation (\ref{EQxi-1}),
$\left\langle \xi_{PQ}\right\rangle =0$.

From the relation $\rho_{QP}\left(t\right)=\mathrm{Tr}\left[\rho\left(t_{0}\right)\sigma_{PQ}\left(t\right)\right]$
one immediately deduces the evolution equation for the density matrix

\begin{eqnarray*}
\partial_{t}\rho_{QP} & \approx & -\left(\Gamma_{PQ}+\mathrm{i}\Delta_{PQ}\right)\rho_{QP}\left(t\right)+\delta_{PQ}\sum_{P<M}\gamma_{MP}\rho_{MM}\left(t\right)
\end{eqnarray*}
In particular, for coherences $\left(P\neq Q\right)$, we obtain
\begin{eqnarray*}
\rho_{QP}\left(t\right) & \approx & e^{-\left(\Gamma_{PQ}+\mathrm{i}\Delta_{PQ}\right)\left(t-t_{0}\right)}\rho_{QP}\left(t_{0}\right).
\end{eqnarray*}

\section{Atomic data\label{APPAtomicData}}

In order to calculate the rates of spontaneous emission for levels
$10s_{1/2}$ and $10p_{1/2,3/2}$, we need energies and transition
dipole moments involving $s$, $p$ and $d$ lower levels. Regarding
energies, we take experimental values from the NIST database \citep{NIST_ASD}.
Transition dipole moments are calculated using the Cowan codes \citep{cowan1981}.

The vector associated with the dipole operator is expressed as irreducible
tensors $\hat{d}_{q}$ ($q=0,\pm1$), such that $\hat{d}_{0}=\hat{d}_{z}$
and $\hat{d}_{\pm1}=\mp(\hat{d}_{x}\pm i\hat{d}_{y})\sqrt{2}$. Their
matrix elements in the coupled atomic basis $\{|n\ell_{j}m_{j}\rangle\}$
read 
\begin{eqnarray}
\left\langle n'\ell'_{j'}m_{j}'\right|\hat{d}_{q}\left|n\ell_{j}m_{j}\right\rangle  & = & e(-1)^{j+\ell'+s}\sqrt{(2j+1)}\times\left\{ \begin{array}{ccc}
\ell & s & j\\
j' & 1 & \ell'
\end{array}\right\} \left\langle n'\ell'\right\Vert \hat{\mathbf{r}}\left\Vert n\ell\right\rangle C_{jm_{j}1q}^{j'm_{j}'}\label{eq:dq-1}
\end{eqnarray}
where $e$ is the absolute value of the electron charge, $\left\{ \begin{array}{ccc}
a & b & c\\
d & e & f
\end{array}\right\} $ is a Wigner 6j symbol, and $C_{a\alpha b\beta}^{c\gamma}$ a Clebsch-Gordan
coefficient \citep{varshalovich1988}. The quantity $\langle n'\ell'\Vert\hat{\mathbf{r}}\Vert n\ell\rangle$
is the reduced matrix element of the position operator of the outermost
electron. In our calculations, it is supposed to be independent from
$j$ and $j'$.

\begin{table}
\begin{centering}
\begin{tabular}{rrrr}
\hline 
$n'$  & $\langle n'p\Vert\hat{\mathbf{r}}\Vert10s\rangle$  & $\langle n's\Vert\hat{\mathbf{r}}\Vert10p\rangle$  & $\langle n'd\Vert\hat{\mathbf{r}}\Vert10p\rangle$\tabularnewline
\hline 
3  & 0.1458095  & 0.0205458  & -0.0159734 \tabularnewline
4  & 0.3574532  & 0.0914227  & -0.0856747 \tabularnewline
5  & 0.7398834  & 0.2407079  & -0.2638617 \tabularnewline
6  & 1.4886102  & 0.5381275  & -0.6944033 \tabularnewline
7  & 3.2367546  & 1.1745454  & -1.8591371 \tabularnewline
8  & 9.1040991  & 2.8038037  & -6.1298872 \tabularnewline
9  & 71.1485790  & 8.9586411  & 160.0011787 \tabularnewline
10  &  & 93.2001425  & \tabularnewline
\hline 
\end{tabular}
\par\end{centering}
\caption{\label{tab:r} Reduced matrix elements $\langle n'\ell'\Vert\hat{\mathbf{r}}\Vert n\ell\rangle$
in units of $ea_{0}$ ($e$ is the absolute value of the electron
charge and $a_{0}$ is the Bohr radius), for $n\ell=10s$ and $10p$.}
\end{table}
Table \ref{tab:r} contains the quantities $\langle n'\ell'\Vert\hat{\mathbf{r}}\Vert10s\rangle$
and $\langle n'\ell'\Vert\hat{\mathbf{r}}\Vert10p\rangle$ relevant
for our calculation. They give radiative lifetimes of 0.855, 8.58
and 8.56 $\mu s$ for $10s$, $10p_{1/2}$ and $10p_{3/2}$ respectively,
which are in correct agreement with the values reported in \citep{theodosiou1984}.


\begin{thebibliography}{99}
\bibitem{DecoupledBasis}We recall the relations of the coupled basis
states with the decoupled basis states : $\begin{array}{lcc}
\left|j=\frac{3}{2},m_{j}=-\frac{3}{2}\right\rangle  & = & \left|l=1,m_{l}=-1;s=\frac{1}{2},m_{s}=-\frac{1}{2}\right\rangle \\
\left|j=\frac{3}{2},m_{j}=-\frac{1}{2}\right\rangle  & = & \sqrt{\frac{2}{3}}\left|l=1,m_{l}=0;s=\frac{1}{2},m_{s}=-\frac{1}{2}\right\rangle \\
 &  & +\frac{1}{\sqrt{3}}\left|l=1,m_{l}=-1;s=\frac{1}{2},m_{s}=\frac{1}{2}\right\rangle \\
\left|j=\frac{3}{2},m_{j}=\frac{1}{2}\right\rangle  & = & \frac{1}{\sqrt{3}}\left|l=1,m_{l}=1;s=\frac{1}{2},m_{s}=-\frac{1}{2}\right\rangle \\
 &  & +\sqrt{\frac{2}{3}}\left|l=1,m_{l}=0;s=\frac{1}{2},m_{s}=\frac{1}{2}\right\rangle \\
\left|j=\frac{3}{2},m_{j}=\frac{3}{2}\right\rangle  & = & \left|l=1,m_{l}=1;s=\frac{1}{2},m_{s}=\frac{1}{2}\right\rangle \\
\left|j=\frac{1}{2},m_{j}=-\frac{1}{2}\right\rangle  & = & \frac{1}{\sqrt{3}}\left|l=1,m_{l}=0;s=\frac{1}{2},m_{s}=-\frac{1}{2}\right\rangle \\
 &  & -\sqrt{\frac{2}{3}}\left|l=1,m_{l}=-1;s=\frac{1}{2},m_{s}=\frac{1}{2}\right\rangle \\
\left|j=\frac{1}{2},m_{j}=\frac{1}{2}\right\rangle  & = & \sqrt{\frac{2}{3}}\left|l=1,m_{l}=1;s=\frac{1}{2},m_{s}=-\frac{1}{2}\right\rangle \\
 &  & -\frac{1}{\sqrt{3}}\left|l=1,m_{l}=0;s=\frac{1}{2},m_{s}=\frac{1}{2}\right\rangle .
\end{array}$
\end{thebibliography}
\end{document}